\documentclass[
reprint,
groupedaddress,
amsmath,amssymb,
prb,
floatfix,
superscriptaddress
]{revtex4-1}
\usepackage{xcolor}

\usepackage{amsmath}
\usepackage{bm}
\usepackage{amsfonts}
\usepackage{graphicx}
\usepackage{amssymb}
\usepackage{sidecap}
\usepackage[caption=false]{subfig}
\usepackage{hyperref}
\usepackage{cleveref}
\usepackage{siunitx}
\usepackage[T1]{fontenc}
\usepackage{libertine}
\usepackage{lineno}

\begin{document}

\title{Ferroelastic toughening: can it solve the mechanics challenges of solid electrolytes?}
\author{Anton Van der Ven}
\email{avdv@ucsb.edu}
\affiliation{Materials Department, University of California Santa Barbara}
\author{Robert M. McMeeking}
\email{rmcm@ucsb.edu}
\affiliation{Materials Department, University of California Santa Barbara}
\affiliation{Department of Mechanical Engineering, University of California Santa Barbara}
\author{Rapha\"ele J. Cl\'ement}
\email{rclement@ucsb.edu}
\affiliation{Materials Department, University of California Santa Barbara}
\affiliation{Materials Research Laboratory, University of California Santa Barbara}
\author{Krishna Garikipati}
\email{krishna@umich.edu}
\affiliation{Department of Mechanical Engineering, University of Michigan Ann Arbor}

\date{\today}

\begin{abstract}

The most promising solid electrolytes for all-solid-state Li batteries are oxide and sulfide ceramics. 
Current ceramic solid electrolytes are brittle and lack the toughness to withstand the mechanical stresses of repeated charge and discharge cycles.
Solid electrolytes are susceptible to crack propagation due to dendrite growth from Li metal anodes and to debonding processes at the cathode/electrolyte interface due to cyclic variations in the cathode lattice parameters. 
In this perspective, we argue that solutions to the mechanics challenges of all-solid-state batteries can be borrowed from the aerospace industry, which successfully overcame similar hurdles in the development of thermal barrier coatings of superalloy turbine blades. 
Their solution was to exploit ferroelastic and transformation toughening mechanisms to develop ceramics that can withstand cyclic stresses due to large variations in temperature. 
This perspective describes fundamental materials design principles with which to search for solid electrolytes that are ferroelastically toughened.

\end{abstract}

\maketitle
\section{Introduction}\label{sec:intro}

Li-ion batteries continue to be the preferred power source of portable devices and electrified transportation. 
Significant increases in energy density and safety are anticipated with the introduction of all-solid-state batteries enabled by solid electrolytes.\cite{manthiram2017lithium,zhang2018new,famprikis2019fundamentals,kim2021solid} 
Solid electrolytes replace flammable liquid electrolytes and may serve as a mechanical barrier against dendrite formation and growth, thereby opening the door to the use of metallic Li anodes and to significant increases in energy density.
They also offer more flexibility in terms of battery design compared to their liquid counterparts and allow more compact, weight-saving, and cost-saving configurations.\cite{jung2019solid,famprikis2019fundamentals} 

In spite of the tremendous promise of solid electrolytes, significant challenges persist to make them commercially viable. 
Solid electrolytes must have ion conductivities that approach those of liquid electrolytes. 
Progress has been made on this front with the discovery of superionically-conducting sulfide compounds that exhibit Li-ion conductivities up to ~10-15 mS/cm.\cite{kamaya2011lithium,kato2016high,zhang2018new} 
Sulfides, however, tend to decompose when placed in contact with common cathodes and anodes of commercial Li-ion batteries or when exposed to moisture in ambient air.\cite{kim2021solid}
Oxide solid electrolytes such as those derived from Li$_7$La$_3$Zr$_2$O$_{12}$ (LLZO)\cite{murugan2007fast}, in contrast, are more chemically and electrochemically stable,\cite{kim2021solid} but have lower ionic conductivities (on the order 1 mS/cm) than the best sulfide electrolytes \cite{kamaya2011lithium}. 
Nevertheless, fundamental insights into Li-ion diffusion mechanisms within those structures\cite{van2013understanding,wang2015design,he2017origin,siegel2021establishing,jun2022lithium} suggest that further improvements in conductivity can be achieved through rational materials design.  

A major impediment to the widespread implementation of all-solid-state batteries that has so far received less attention is the poor mechanical properties of typical solid electrolytes.\cite{zhang2020review}
For instance, oxides are brittle and recent studies have revealed that dendrites are in fact able to penetrate them, both along grain boundaries as well as through single crystal grains.\cite{ren2015direct,porz2017mechanism,cheng2017intergranular,han2019high,krauskopf2019toward,kazyak2020li,liu2021local} 
The mechanisms by which this occur remain poorly understood and constitute an active area of research.
Another major challenge is the maintenance of coherency between the solid electrolyte layer and the electrodes during cycling, especially at the cathode-electrolyte interface.\cite{zhang2020review}
While the solid electrolyte maintains a constant Li concentration and therefore constant lattice parameters during a charge-discharge cycle, the cathode undergoes significant changes in Li concentration that are often accompanied by large variations in lattice parameters.\cite{radin2017narrowing,radin2017role} 
The cyclic mismatch in lattice parameters as the cathode structure evolves during charge and discharge leads to mechanical degradation at the electrode-electrolyte interface and ultimately to the loss of coherency and ionic contact.\cite{zhang2020review} 

Here, we argue that, to solve the coherency problem in all-solid-state batteries, and to improve overall mechanical properties of solid electrolytes, inspiration can be drawn from the successful development of thermal barrier coatings of jet engine turbine blades.\cite{clarke2003materials,pollock2012multifunctional} 
This technology has similar two-phase coherency challenges due to the cyclic temperature fluctuations that are imposed during service on the metal-ceramic heterostructure coatings of the superalloy blade.
The constituent phases of the heterostructure have different coefficients of thermal expansion and their lattice parameters expand and contract with different rates upon heating and cooling, leading to large coherency stresses. 
The problem was in part solved with the use of Y-stabilized ZrO$_2$ (YSZ) coatings that are transformation \cite{evans2001mechanisms} and ferroelastically\cite{virkar1986ferroelastic,mercer2007ferroelastic} toughened. 
The transformation of a high temperature cubic phase of YSZ to a twinned microstructure consisting of multiple tetragonally and monoclinically distorted variants produces a toughness and a resistance to crack propagation that far exceeds that of typical ceramics \cite{clarke2003materials}. 
In this contribution, we describe how similar principles may be applied to solve the coherency and dendrite propagation challenges of all-solid-state batteries by increasing the toughness of ceramic solid electrolytes. 
We focus in particular on the exploitation of ferroelastic toughening mechanisms that can be achieved with highly twinned microstructures, enumerate fundamental design principles to guide materials selection, and highlight specific materials classes of interest for further exploration. The materials design principles identified herein are applicable to Li-based batteries and beyond, including emerging Na-based systems that have received significant attention as more sustainable alternatives to the current Li technology and that exhibit similar physicochemical properties as their Li counterparts.

\section{Ferroelastic toughening}

Following the example of ferroelastically toughened YSZ,\cite{virkar1986ferroelastic,mercer2007ferroelastic} we contend that solid electrolytes with carefully tuned twinned microstructures will overcome many of the mechanics challenges facing the widespread implementation of all-solid-state batteries. 
The twinned microstructures can be generated by means of a group/subgroup structural transformation from a high symmetry crystal structure, stable at elevated temperature, to a lower symmetry distorted phase. 
A slight symmetry breaking strain is necessary to ensure that there are multiple variants that can coexist in a twinned microstructure.
Such a twinned microstructure of coexisting lower-symmetry variants can be engineered to have a toughness to resist crack propagation and to withstand the cyclic stresses imposed by the electrodes that it is coherently attached to.
We illustrate the principle in two dimensions first and then generalize to three dimensions.

\begin{figure}[htbp]
  \centering
  \includegraphics[width=8cm]{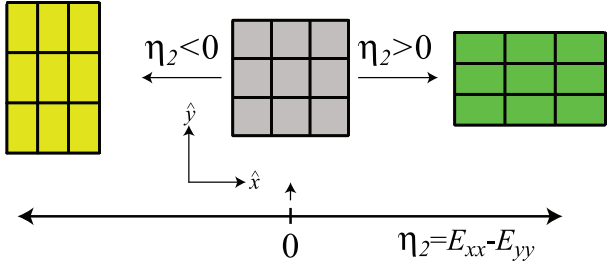} 
   \caption{The strain order parameter, $\eta_2$ represents symmetry-breaking from the square to the rectangular lattices.}
  \label{fig:square_rectangle}
\end{figure}

Consider the simple two-dimensional square to rectangle structural phase transformation. 
A hypothetical compound may be stable in the square lattice at high temperature, but transform to a rectangularly distorted lattice at low temperature. 
The symmetry of the square lattice ensures that there are two rectangular variants that can emerge at low temperature. 
One can be generated by elongating the square lattice along the $\hat{x}$ axis, while the other can form when elongating along the $\hat{y}$ axis as illustrated in Figure \ref{fig:square_rectangle}. 
The square to rectangle phase transformation can be tracked with symmetry adapted linear combinations of the Cartesian strains with the strain order parameters $\eta_{1} = 1/\sqrt{2}(E_{xx} + E_{yy})$, $\eta_{2}=1/\sqrt{2}(E_{xx} - E_{yy})$ and $\eta_{3}=\sqrt{2}E_{xy}$.\cite{jacobs1985solitons}
While, in the limit of infinitesimal strains, $\eta_1$ measures symmetry preserving volumetric deformations and $\eta_{3}$ is a measure of a shear strain, $\eta_2$ is of particular interest for the square to rectangle phase transformation, as it can distinguish between the square reference phase and the two rectangular variants (Figure \ref{fig:square_rectangle}). 
The square lattice corresponds to $\eta_2 = 0$, while the rectangular variant stretched along the $\hat{x}$-axis ($\hat{y}$-axis) has a positive (negative) value for $\eta_2$.
The symmetry of the rectangular distortion is a subgroup of the square lattice.

\begin{figure}[htbp]
  \centering
  \includegraphics[width=6cm]{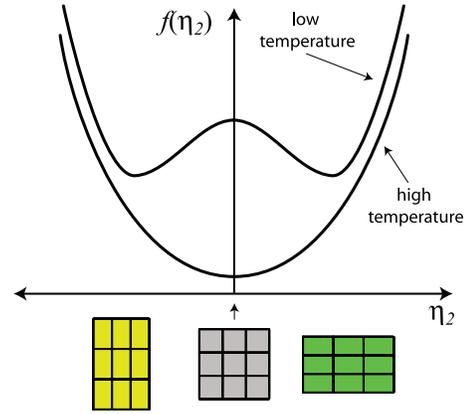} 
   \caption{Slices through a schematic free energy surface, $f(\eta_2)$, at low and high temperatures illustrate the thermodynamics underlying the symmetry-breaking transformation from a single, square lattice at high temperature to two rectangular variants at low temperature corresponding to the pair of local minima.}
  \label{fig:free_energy}
\end{figure}

A structural transformation from the square lattice to the rectangular lattice, with the group/subgroup relation between their symmetries, can emerge if the compound has free energy surfaces as a function of temperature and the order parameter $\eta_2$ similar to those illustrated in Figure \ref{fig:free_energy}. 
At high temperature, the free energy has a minimum at $\eta_2=0$, signifying the stability of the square lattice. 
At low temperature, the square lattice becomes unstable as manifested by the negative curvature of the free energy at $\eta_2=0$ and two symmetrically equivalent free energy wells emerge at finite values of $\eta_2$, revealing a more stable rectangular lattice. 
A square lattice that is quenched from high temperature will become elastically unstable and martensitically collapse into one of the two symmetrically equivalent rectangular variants. 

\begin{figure}[htbp]
  \centering
  \includegraphics[width=8cm]{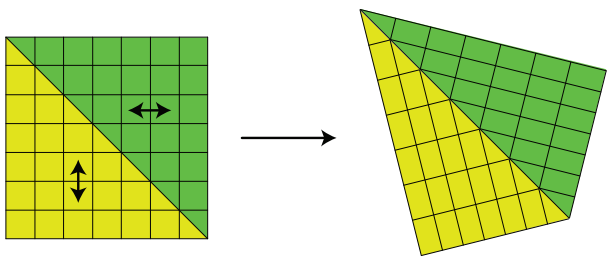} 
   \caption{The square to rectangle martensitic transformation with an atomically sharp twin boundary between rectangular variants.}
  \label{fig:twin_boundary}
\end{figure}

\begin{figure}[htbp]
  \centering
  \includegraphics[width=8cm]{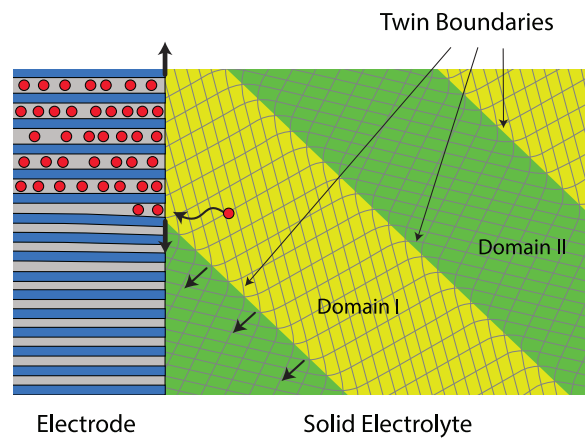} 
   \caption{The proposed transformation toughening mechanism. The schematically indicated intercalation of lithium atoms into the electrode causes an expansion of interlayer spacing on the left. The corresponding strain can be accommodated by expansion of favorably aligned domains in the electrolyte to the right. The rectangular variant domains and the diffuse twin boundaries between them are the result of a square to rectangle martensitic transformation computed with a gradient theory of nonconvex elasticity  at finite strain.\cite{rudraraju2016mechanochemical}}
  \label{fig:coherent_interface}
\end{figure}

Due to spatial inhomogeneities and constraints from boundary conditions, some regions of the unstable square phase will collapse into the variant elongated along the $\hat{x}$-axis, while others will collapse into the variant elongated along the $\hat{y}$-axis. 
If the change in dimensions from the square lattice to the rectangular lattice is not too large, the boundaries between different variants can remain coherent and a fully coherent twinned microstructure will emerge. 
Different variants thus coexist along coherent twin boundaries, where the strain order parameter varies from the equilibrium value of one variant to that of the other variant. 
The twin boundary can be atomically sharp as illustrated in Figure \ref{fig:twin_boundary}, or it can be gradual, characterized by a diffuse variation of the strain order parameter $\eta_2$ between its equilibrium values in the two neighboring variants. The latter case appears in Figure \ref{fig:coherent_interface}, where the coexisting variants and the diffuse twin boundaries between them are the result of a state computed from a gradient theory of nonconvex elasticity at finite strains.\cite{rudraraju2016mechanochemical}

Twin boundaries can be mobile. 
A tensile load on a twinned microstructure will favor the variant whose elongated axis is most closely aligned with the direction of the imposed load.  
The favorably oriented variant will enlarge at the expense of the less favorably oriented variant through the migration of the twin boundary, thereby leading to a macroscopic strain. 
This is referred to as ferroelastic toughening and is one of the mechanisms responsible for the toughening of YSZ.\cite{virkar1986ferroelastic,mercer2007ferroelastic}
We suggest that this phenomenon can be exploited to increase the fracture toughness of solid electrolytes and to maintain coherency between an electrode and a solid electrolyte. 

\begin{figure}[htbp]
    \centering
    \includegraphics[width=6cm]{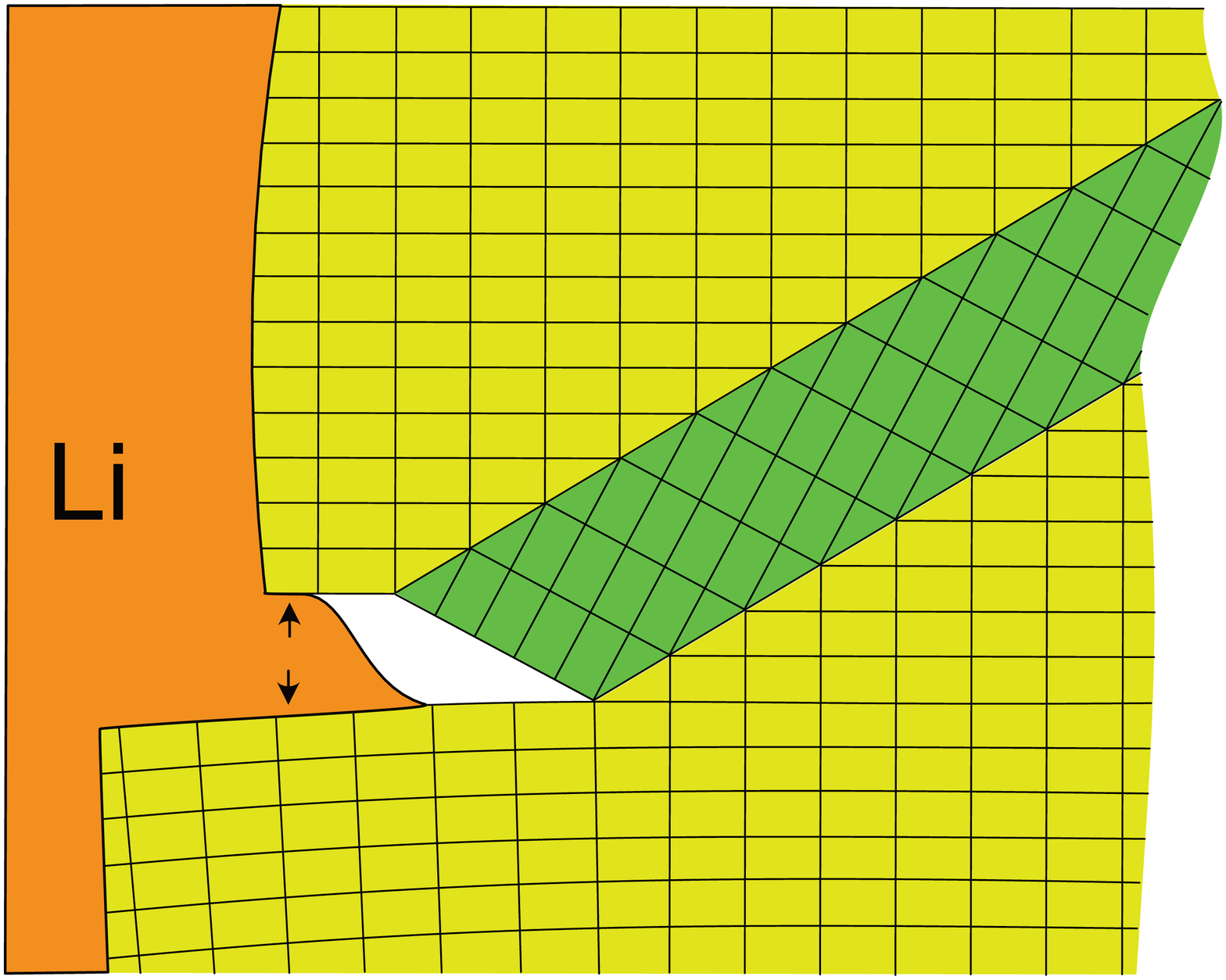}
    \caption{Ferroelastic toughening mechanisms facilitate crack tip blunting and thereby relieve high stress intensities at crack tips as Li dendrites grow along cracks.}
    \label{fig:Crack_tip_blunting}
\end{figure}

Ferroelastic toughening mechanisms will help alleviate cyclic stress fluctuations at the electrode/electrolyte interface and thereby suppress debonding between the electrode and electrolyte (Figure \ref{fig:coherent_interface}).
Li insertion and extraction into/from commercial layered intercalation cathodes, such as Li$_x$CoO$_2$ or Li$_x$(Ni$_{1-y-z}$Mn$_{y}$Co$_{z}$)O$_2$, leads to cyclic changes in the interlayer spacing and anisotropic volume changes during charge and discharge. \cite{radin2017narrowing}  
Many layered Li and Na intercalation compounds also under go gliding phase transformations \cite{amatucci1996coo2,van1998first,chen2002staging,vinckeviciute2016stacking,kaufman2019understanding} that induce shape changes of the electrode particles.\cite{radin2017role}
A stabilizing mechanism is proposed, whereby a ferroelastically-toughened solid electrolyte coherently joined to such electrode materials, as schematically illustrated in Figure \ref{fig:coherent_interface}, is able to minimize large elastic stresses along the coherent interface by locally expanding or contracting through the migration of twin boundaries.
The solid electrolyte then remains coherently attached by accommodating the strain of the cathode due to (de)lithiation by expanding/contracting favorably oriented domains through twin boundary migration as illustrated in Figure \ref{fig:coherent_interface}.
Ferroelastic toughening mechanisms should also be effective in suppressing crack propagation due to Li dendrites by blunting sharp crack tips as schematically illustrated in Figure \ref{fig:Crack_tip_blunting}
While the mechanisms that spur Li dendrite growth through solid electrolytes remain unclear, fracture mechanics studies \cite{klinsmann2019dendritic,shishvan2020dendrites,shishvan2020growth} show that Li dendrites can produce high stress intensities at crack tips. 
These can be alleviated by the pseudoplasticity due to ferroelastic behavior to increase the fracture toughness of the solid electrolyte. Furthermore, energy dissipation by ferroelastic transformation toughening is another candidate mechanism that could alter the thermodynamic driving forces for dendrite growth.

\begin{figure}[htbp]
    \centering
    \includegraphics[width=8cm]{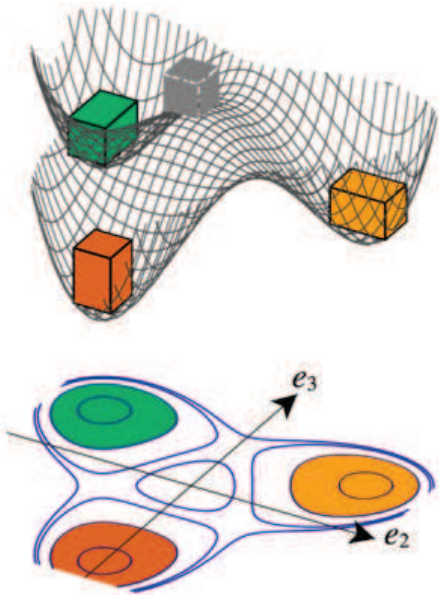}
    \caption{The cubic to tetragonal transformation with three resulting variants, and the corresponding free energy surface.}
    \label{fig:3Dcubic-tetragonal}
\end{figure}

The principles of ferroelastic toughening can be extended to three dimensions. 
The canonical example is the exploitation of the cubic to tetragonal structural phase transition to generate a three-dimensional twinned microstructure that accommodates macroscopic strains through twin boundary migration. 
Here again, strain order parameters can be introduced as symmetry adapted linear combinations of the six Green-Lagrange strain components, traditionally computed with respect to a Cartesian coordinate system aligned with the cubic axes, to track the extent and nature of distortions that accompany the group/subgroup transformation. 
Of the six resulting strain order parameters,\cite{barsch1984twin,thomas2017exploration} two, defined as $e_2=(E_{xx}-E_{yy})/\sqrt{2}$ and $e_3=(2E_{zz}-E_{xx}-E_{yy})/\sqrt{6}$, are able to distinguish the high symmetry cubic phase from the three tetragonal variants that can form by stretching the $\hat{z}$, $\hat{y}$ or $\hat{x}$ axes. This is illustrated in Figure \ref{fig:3Dcubic-tetragonal}
 
Similar to the two dimensional square to rectangle transformation, a free energy can be defined in terms of the two strain order parameters that has a minimum at $e_2 = e_3 = 0$ at high temperature, corresponding to the cubic phase. 
This surface changes shape to having a local maximum at $e_2 = e_3 = 0$ and three symmetrically equivalent minima at finite strains separated by angles of $2\pi/3$ in the $e_2$-$e_3$ plane as illustrated schematically in Figure \ref{fig:3Dcubic-tetragonal}. 
A quench of the cubic phase will then form a twinned microstructure consisting of a mixture of different orientational variants of the tetragonal phase, separated by twin boundaries. 

A variety of multi-scale theories have been developed to rigorously model ferroelastic and transformation toughening.\cite{mcmeeking1982mechanics,budiansky1983continuum,rudraraju2014three,rudraraju2016mechanochemical}
These will prove invaluable to establish optimal materials properties that a solid electrolyte, in combination with a particular pair of electrodes, should possess.  
First, the formation of a twinned microstructure and its response to mechanical loads can be modeled with a continuum phase-field theory.\cite{barsch1984twin, rudraraju2016mechanochemical}
The strain order parameters serve as the phase-field variables to distinguish between the different twinned variants with the free energy of the inhomogenous solid depending not only on the local strain variables, but also on gradients of strain variables. 
%
Furthermore, numerical approaches to solve mechanics problems that account for strain gradients in the free energy description have recently been developed and implemented. \cite{rudraraju2014three,rudraraju2016mechanochemical, sagiyama2016unconditionally,wang2016three,teichert2017variational,sagiyama2017numerical}
Finally, the connection between continuum descriptions of microstructure evolution and the electronic properties of a material can be made with first-principles statistical mechanics.\cite{van2018first,van2020rechargeable}
These modeling tools will guide the establishment of materials design principles and materials chemistries that overcome many of the challenges posed by the brittle mechanical properties of ceramic solid electrolytes.

We emphasize that reversible and non-destructive deformation mechanisms are desired since the solid electrolyte must undergo cyclic deformations and maintain structural integrity. 
This will require a sufficient number of twin boundaries that can migrate with minimal frictional resistance. 
It also means that other mechanisms of deformation, such as dislocation nucleation and glide, deformation twinning and trans/inter granular crack propagation must be more energetically costly than twin boundary migration. 
The dominance of different deformation mechanisms can be assessed with calculations of generalized stacking fault energies, cleavage energies and barriers of twin boundary migration.\cite{enrique2017decohesion,liu2017atomistic,goiri2021multishifter,san2022uncovering}
Facile twin boundary migration is also desirable to minimize hysteresis due to the electrochemical overpotentials that are required to overcome the mechanical work of ferroelastic deformation.\cite{van2022hysteresis}

\section{Materials requirements and challenges}

The search for solid electrolytes has primarily been guided by design principles that seek to increase ionic conductivity, electrochemical stability and elastic stiffness.\cite{manthiram2017lithium,zhang2018new,famprikis2019fundamentals,kim2021solid} 
The design of solid electrolytes that are ferroelastically toughened hinges on additional materials properties, namely the ability of a solid electrolyte material to form twinned microstructures with highly mobile twin boundaries. 
As with yttria stabilized zirconia, this can be achieved if the solid electrolyte undergoes a phase transformation upon cooling that is accompanied by a symmetry-breaking homogeneous strain of a high symmetry parent crystal structure.
The candidate compound should have free energy surfaces similar to those of Figure \ref{fig:free_energy}, which predict a high symmetry phase at elevated temperature that becomes unstable at low temperature with respect to a symmetry-breaking homogeneous strain. 

\begin{figure}[htbp]
  \centering
  \includegraphics[width=8cm]{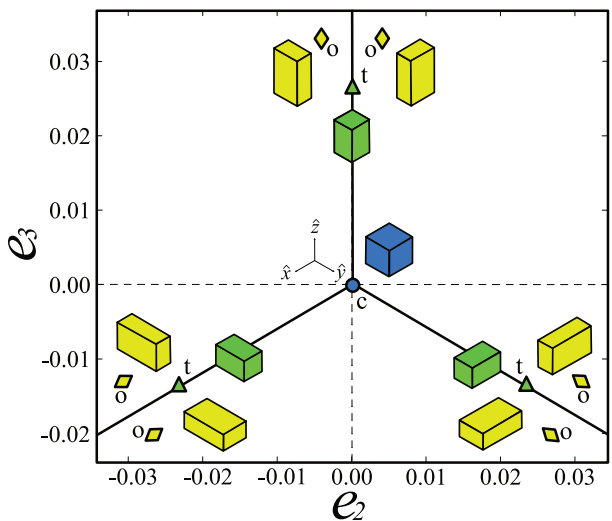} 
   \caption{A hierarchy of symmetry breaking distortions can be visualized as a function of strain order parameters such as $e_2$ and $e_3$ (see text). The strain order parameters $e_2$ and $e_3$ describe tetragonal (t) and orthorhombic (o) distortions of a cubic (c) reference crystal (shown at the origin). Adapted from Ref.\cite{behara2022ferroelectric}}
  \label{fig:strain_map}
\end{figure}

The existence of a high symmetry parent crystal ensures that multiple orientational variants of a lower symmetry phase can coexist in a twinned microstructure. 
Figure \ref{fig:strain_map} illustrates relationships between high symmetry parent lattices and the multiple orientational variants of lower symmetry lattices whose formation is accompanied by a small symmetry-breaking strain. 
The relationships are most clearly revealed when mapped as a function of symmetry adapted strain order parameters such as $e_2$ and $e_3$ defined earlier.\cite{thomas2017exploration,bechtel2018octahedral,behara2022ferroelectric} 
As is clear in Figure \ref{fig:strain_map}, a cubic parent (blue) can form three orientational variants of a tetragonal child (green), while a tetragonal parent can form two orientational variants of an orthorhombic child (yellow). 
An orthorhombic parent in turn can transform into four orientational variants of a monoclinic child.\cite{behara2022ferroelectric} 
The orthorhombic to monoclinic distortion requires a shear strain in addition to $e_2$ and $e_3$. 
Another common symmetry breaking distortion is from a cubic parent to one of four orientational variants of a rhombohedral unit cell, which can be described with the three shear strains $e_4=\sqrt{2}E_{yz}$, $e_{5}=\sqrt{2}E_{xz}$ and $e_{6}=\sqrt{2}E_{xy}$.\cite{thomas2017exploration,bechtel2018octahedral}




A wide variety of solid electrolytes with diverse crystal structures are actively being investigated for Li-ion and Na-ion batteries.\cite{manthiram2017lithium,zheng2018review} 
Many, such as NASICONs (e.g., Na$_3$Zr$_2$Si$_2$PO$_12$)\cite{hong1976crystal,goodenough1976fast}, LISICONs (e.g., $\gamma$-Li$_3$PO$_4$)\cite{hu1977ionic,shannon1977new,hong1978crystal,masquelier1995chemistry}, thio-LISICONs\cite{kanno2001lithium,adams2012structural}, $\beta$-alumina,\cite{yao1967ion} argyrodites,\cite{deiseroth2008li6ps5x,rao2011studies} and garnets (e.g., Li$_7$La$_3$Zr$_2$O$_12$),\cite{klenk2015local,xia2016ionic} have complex crystal structures with unit cells containing a large number of atoms. 
Most of those structures cannot be viewed as lower symmetry distortions of a real or hypothetical higher symmetry parent crystal structure to enable the coexistence of orientational variants in a twinned microstructure. 
Chemical modifications of existing solid electrolyte phases or the use of new classes of crystal structures will, therefore, be necessary to achieve the crystallographic features that are conducive to the formation of twinned microstructures.

The symmetry breaking to produce twinned microstructures in ceramic solid electrolytes can be realized in one of two ways. 
One is by means of an order-disorder reaction whereby the mobile cations and vacancies adopt an ordered arrangement that is accompanied by a change in the symmetry of the crystal upon cooling.\cite{natarajan2017symmetry,van2018first} 
A second way is as a result of a weak martensitic structural transition \cite{bhattacharya2004crystal} driven by low temperature dynamical instabilities.\cite{thomas2013finite,thomas2014elastic,bechtel2019finite} 
We describe these phenomena in more detail and discuss their pros and cons. 

\begin{figure}[htbp]
  \centering
  \includegraphics[width=9cm]{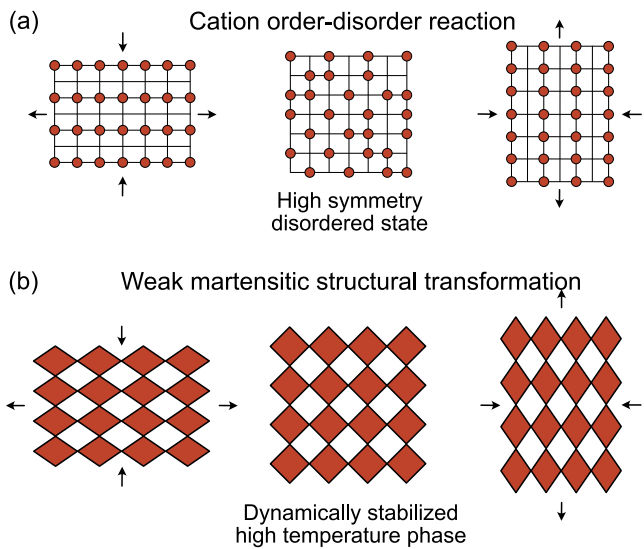} 
   \caption{Symmetry breaking in a crystalline solid electrolyte can be realized through an order-disorder reactions among the mobile cations and vacancies (a) or through a weak martensitic structural transformation. A high symmetry phase can arise due to cation disorder (a) or to large anharmonic vibrational excitations (b). }
  \label{fig:order_disorder}
\end{figure}

Many solid electrolytes undergo an ordering reaction among their mobile cations and vacancies upon cooling from high temperatures.\cite{boilot1979phase,goodenough1984review,awaka2009synthesis,jansen1992synthesis,tang2017order}
In general, the cation-vacancy ordering at low temperature breaks the translational symmetry of the host crystal due to the formation of a super lattice and can appear as one of multiple orientational variants.\cite{natarajan2017symmetry} 
This is schematically illustrated in Figure \ref{fig:order_disorder}(a).
It is this phenomenon that is responsible for the cubic to tetragonal transition of Li$_7$La$_3$Zr$_2$O$_{12}$ (LLZO) with decreasing temperature.\cite{awaka2009synthesis}
The symmetry breaking due to ion-vacancy ordering will cause a symmetry breaking strain of the host unit cell that can potentially be exploited to achieve ferroelastic toughening (Figure \ref{fig:order_disorder}(a)).
Unfortunately, order-disorder reactions in solid electrolytes can have a deleterious effect on the ionic conductivity.\cite{awaka2009synthesis,wolfenstine2012high}
Tetragonal LLZO with ordered Li ions, for example, has a conductivity that is two orders of magnitude lower than that of cubic LLZO in which the Li ions are disordered \cite{wolfenstine2012high}.
The locking in of mobile ions on sublattice sites of an ordered phase is well known to hamper ion mobility\cite{briant1980ionic,van2001first,awaka2009synthesis} and is typically avoided.

The migration of twin boundaries separating orientational variants of an ordered phase is also likely to be sluggish as it will require thermal activation to rearrange Li ions and vacancies through diffusive hops.
(It should be noted, though, that there are examples where the migration of twin boundaries and anti-phase boundaries separating different variants of a particular cation-vacancy ordering through cation migration can be very facile.\cite{kaufman2021antiphase,kaufman2022cation})
The twin boundaries that separate different orientational variants of a Li-vacancy ordered phase also appear to impede Li-ion conduction.\cite{wolfenstine2012high}
For example, the fraction to the total resistance due to boundary effects (e.g. grain boundaries and other internal boundaries) is 10-20 $\%$ in cubic LLZO, while it constitutes 80-90$\%$ in tetragonal LLZO having a similar density and grain size.
The difference has been attributed to the presence of twin boundaries separating differently oriented Li-orderings in tetragonal LLZO.\cite{wolfenstine2012high}
These considerations suggest that ferroelastic toughening mechanisms are unlikely to be achieved by exploiting the symmetry breaking due to order-disorder reactions. 

Symmetry breaking can also arise from a dynamical instability that causes a weak martensitic structural transformation (Figure \ref{fig:order_disorder}(b)). 
It is common that a high symmetry crystal structure is elastically and/or dynamically unstable at low temperature with respect to a symmetry-lowering distortion, only emerging at elevated temperature due to large anharmonic vibrational excitations.\cite{bhattacharya2008mechanical,thomas2013finite,thomas2014elastic, bechtel2019finite, radin2020order} 
These phases have free energy surfaces similar to that of Figure \ref{fig:free_energy} and a quench of the high symmetry phase can result in twinned microstructures of coexisting orientational variants of a lower symmetry polymorph.
Twinned microstructures derived from a weak martensitic structural transformation are likely preferable over those that emerge from order-disorder transitions for solid electrolyte applications. 
Indeed, the twin boundaries generated by a weak martensitic transformation are expected to be more mobile as they do not require diffusive hops. 
A fruitful search direction for ferroelastically toughened solid electrolytes is therefore likely to be found among phases that emerge from high symmetry phases through a weak martensitic structural transformation. 

For many of the candidate oxide and sulfide solid electrolytes mentioned earlier, there is no obvious mechanism with which symmetry breaking distortions can be achieved through a weak martensitic transformation. 
An important exception, however, is the perovskite crystal structure, which is well known to have the crystallographic flexibility to enable a sequence of symmetry lowering structural transformations that can produce twinned microstructures with mobile twin boundaries among lower symmetry polymorphs.\cite{glazer1972classification,howard1998group,carpenter2009symmetry} 

The simple perovskite crystal structure, with stoichiometry ABX$_3$ can adopt 15 polymorphs that are derived from a cubic form through coordinated tilts of the corner sharing BX$_6$ octahedra.\cite{howard1998group} 
Each polymorph is usually accompanied by a small symmetry breaking strain with respect to the cubic parent. 
The stability of lower symmetry variants as well as the transition temperatures of the various structural transformations can be engineered by tuning the Goldschmidt tolerance factor, which is a function of the ionic radii of the constituents, A, B and X, of the perovskite crystal.\cite{goldschmidt1926gesetze,bechtel2018octahedral,lu2021perovskite} 
These can be controlled by modifying cation and anion chemistry. 
Promising solid electrolytes that adopt a perovskite-like crystal structure include Li$_{3x}$La$_{2/3-x}$TiO$_3$ (LLTO) and (Li, Sr)(B,B')O$_3$ where B are cations with a +4 oxidation state (B = Ti, Zr, Hf, Sn, ...) and B' are cations with a +5 oxidation state (B'= Nb, Ta, ...).\cite{lu2021perovskite}
In these compounds, the Li ions share the A sites of the perovskite structure with larger cations such as La and Sr. 
This can lead to cation ordering over the A sites\cite{lu2021perovskite} and thereby cause an additional form of symmetry breaking that is unlikely to be beneficial for twin formation and ferroelastic toughening. 
New chemistries are, therefore, desired that are able to exploit the rich variety of group/subgroup structural transformations of perovskite crystal structures without undergoing order-disorder transitions. 
A promising class of solid electrolytes in this respect are the anti-perovskites, such as Li$_3$OX and Na$_3$OX, where X is a halide (Cl, I, Br).\cite{xia2022antiperovskite,dawson2021anti,deng2022anti} 
These compounds also adopt the perovskite crystal structure but with a reversal of anion and cation positions. 
Similar to perovskites, cubic anti-perovskites also exhibit dynamical instabilities\cite{chen2015anharmonicity} that can result in the same symmetry breaking structural transformations of conventional perovskites. 
While the conductivities of perovskite and anti-perovskite solid electrolytes are still one to two orders of magnitude lower than the best sulfide solid electrolytes based on Li$_{10}$GeP$_2$S$_{12}$ and the argyrodites,\cite{zheng2018review} they remain a competitive and promising class of solid electrolyte materials to meet the stringent chemical, electrochemical and kinetic materials requirements of Li-ion batteries.\cite{lu2021perovskite} 
Their crystallographic flexibility to host twinned microstructures also makes them an appealing class of materials to meet the mechanical challenges of all solid-state batteries.

While perovskite derived compounds are an obvious starting point in the search for ferroelastically toughened solid electrolytes, there are undoubtedly other crystal structure types that can combine high ionic conductivity, electrochemical stability and a susceptibility to group/subgroup structural transformations to enable twinned microstructures. 
The added materials requirements to achieve twinned microstructures conducive to ferroelastic toughening will introduce challenges, though, that need to be overcome.
When a reduction in symmetry arises from a structural distortion, ion diffusion properties may become highly anisotropic. 
In common polycrystalline solid electrolytes, such orientation-dependent ionic conduction at the crystallite level can be problematic as those crystals are typically not co-aligned; this may result in poor intergrain diffusion and a significantly reduced bulk ionic conductivity.
%
Strategies must therefore be identified to achieve high ionic conductivity in low symmetry phases that are derived from a higher symmetry phase. 
While somewhat unexpected, there are examples where symmetry breaking distortions open up diffusion channels or lead to a lowering of migration barriers, as for example in anatase TiO$_2$, where a second-order Jahn-Teller distortion increases the volume of the activated state for Li-ion diffusion.\cite{belak2012kinetics} 
It may also be possible that the twinned microstructure itself may offer new pathways for enhanced ion mobility at the mesoscale, even if the low symmetry phase has a lower ionic conductivity than its high symmetry parent.
For example, diffuse twin boundaries between a pair of symmetrically equivalent tetragonal variants may have more of a cubic character and thereby facilitate Li diffusion, serving as superionic conduction pathways through the bulk material.
Hence, while the optimization of ion transport, electrochemical stability and mechanical properties may seem contraindicated, this is not necessarily the case and a ferroelastically toughened solid electrolyte with exceptional ionic conductivity and electrochemical stability may be waiting to be discovered.

\section{Conclusion}

The stringent mechanical requirements of solid electrolytes for all solid-state batteries are unlikely to be met by the brittle ceramics that are currently being studied. 
The successes in the aerospace industry, where mechanics challenges in the ceramic-metal hetero-structure coatings of turbine blades were overcome by exploiting transformation and ferroelastic toughening mechanisms, provide invaluable guidance in the quest to solve the mechanics problems of all solid-state batteries. 
Overcoming the cyclic mechanical stresses accompanying long-term battery operation require reversible deformation and toughening mechanisms that can be realized by ferroelasticity mediated by facile twin boundary migration. 
This perspective described fundamental materials design principles that may assist the identification and inform the optimization of solid electrolyte materials that are ferroelastically toughened. 
Other toughening mechanisms are also possible, including transformation toughening, which can augment ferroelastic toughening mechanisms if the process can be made reversible as in shape memory alloys such as Nitinol. 








\section{Acknowledgement}
The authors are very grateful for valuable insights and suggestions that originated from extended discussions with Dr. Joseph J. Shiang and Dr. Don M. Lipkin at GE Global Research. 
This material is based upon work supported by the Defense Advanced Research Projects Agency under Contract No. HR001122C0097.

\bibliography{./references.bib}
\end{document}